\pdfoutput=1

\documentclass[11pt,twoside,a4paper,cmspaper,final,collab]{cms-tdr}
\def\svnVersion{f7232a5-D}\def\svnDate{2020/05/12}\def\cmsCernNoTag{CERN-EP-2020-043}\def\cmsCernDate{\today}\def\cmsMessage{"Published in Physics Letters B as \href{http://dx.doi.org/10.1016/j.physletb.2020.135502}{\doi{10.1016/j.physletb.2020.135502}.}"}

\begin{document}\cmsNoteHeader{EXO-19-010}

\hyphenation{had-ron-i-za-tion}
\hyphenation{cal-or-i-me-ter}
\hyphenation{de-vices}
\providecommand{\cmsRight}{right\xspace}
\ifthenelse{\boolean{cms@external}}{\providecommand{\cmsTable}[1]{#1}}{\providecommand{\cmsTable}[1]{\resizebox{\textwidth}{!}{#1}}}

\newlength\cmsTabSkip\setlength{\cmsTabSkip}{1ex}

\newcommand{\intLumiA}{\ensuremath{42\fbinv}}
\newcommand{\intLumiB}{\ensuremath{60\fbinv}}
\newcommand{\intLumi}{\ensuremath{101\fbinv}}
\newcommand{\intLumiRunTwo}{\ensuremath{140\fbinv}}
\newcommand{\backgroundAB}{\ensuremath{47.8_{-2.3}^{+2.7}\stat\pm 8.1\syst}}
\newcommand{\bestexclusionWinoThreeNs}{\ensuremath{884}}
\newcommand{\bestexclusionWinoPTwoNs}{\ensuremath{474}}
\newcommand{\bestexclusionHiggsinoThreeNs}{\ensuremath{750}}
\newcommand{\bestexclusionHiggsinoPZeroFiveNs}{\ensuremath{175}}
\newcommand{\observationAB}{\ensuremath{48}}
\newcommand{\Pchipm}{\PSGcpmDo}
\providecommand{\PSGcmpDo}{{\HepParticle{\PSGc}{1}{\mp}}\Xspace}
\newcommand{\Pchimp}{\PSGcmpDo}
\newcommand{\Pchip}{\PSGcpDo}
\newcommand{\Pchim}{\PSGcmDo}
\newcommand{\Pchiz}{\PSGczDo}
\newcommand{\PchizTwo}{\PSGczDt}
\providecommand{\PSGczDot}{{\HepParticle{\PSGc}{1,2}{0}}\Xspace}
\newcommand{\PchizOneTwo}{\PSGczDot}
\newcommand{\Ztoll}{\PZ{\to}\ell\ell}
\newcommand{\Ztoee}{\PZ{\to}\Pe\Pe}
\newcommand{\Ztomumu}{\PZ{\to}\PGm\PGm}
\newcommand{\Ztotautau}{\PZ{\to}\PGt\PGt}
\newcommand{\Wtolnu}{\PW{\to}\ell\PGn}
\newcommand{\Pveto}{P_{\text{veto}}}
\newcommand{\Poffline}{P_{\text{off}}}
\newcommand{\Ptrigger}{P_{\text{trig}}}
\newcommand{\Phemveto}{P_{\text{HEM}}}
\newcommand{\Pfakeraw}{P_{\text{spurious}}^{\text{raw}}}
\newcommand{\Nctrll}{N_{\text{ctrl}}^{\ell}}
\newcommand{\NctrlBasic}{N_{\text{ctrl}}^{\text{basic}}}
\newcommand{\Nestl}{N_{\text{est}}^{\ell}}
\newcommand{\NestFake}{N^{\text{est}}_{\text{spurious}}}
\newcommand{\Ntp}{N_{\text{T\&P}}}
\newcommand{\NtpVeto}{N_{\text{T\&P}}^{\text{veto}}}
\newcommand{\NSStp}{N_{\text{SS\,T\&P}}}
\newcommand{\NSStpVeto}{N_{\text{SS\,T\&P}}^{\text{veto}}}
\newlength\cmsGapLength
\ifthenelse{\boolean{cms@external}}{\setlength\cmsGapLength{-0.30em}}{\setlength\cmsGapLength{-0.45em}}
\newcommand{\metNoMu}{\ensuremath{p_{\text{T}}^{\text{miss},\,\mu\hspace{\cmsGapLength}/}}\xspace}
\newcommand{\metNoMuVec}{\ensuremath{\vec{p}_{\mathrm{T}}^{\kern1pt\text{miss},\,\mu\hspace{\cmsGapLength}/}}\xspace}
\newcommand{\Ecalo}{\ensuremath{E_{\text{calo}}^{\DR < 0.5}}\xspace}
\newcommand{\dPhiJetMET}{\ensuremath{\abs{\Delta\phi (\text{leading jet} \, \metNoMuVec)}}\xspace}
\newcommand{\nlayers}{\ensuremath{n_{\text{lay}}}}

\cmsNoteHeader{EXO-19-010}
\title{Search for disappearing tracks in proton-proton collisions at $\sqrt{s} = 13\TeV$}

\date{\today}

\abstract{
A search is presented for long-lived charged particles that decay within the volume of the silicon tracker of the CMS experiment.
Such particles can produce events with an isolated track that is missing hits in the outermost layers of the silicon tracker, and is also associated with little energy deposited in the calorimeters and no hits in the muon detectors.
The search for events with this ``disappearing track'' signature is performed in a sample of proton-proton collisions recorded by the CMS experiment at the LHC with a center-of-mass energy of 13\TeV, corresponding to an integrated luminosity of $\intLumi$ recorded in 2017 and 2018.
The observation of $\observationAB$ events is consistent with the estimated background of $\backgroundAB$ events.
Upper limits are set on chargino production in the context of an anomaly-mediated supersymmetry breaking model for purely wino and higgsino neutralino scenarios.
At 95\% confidence level, the first constraint is placed on chargino masses in the higgsino case, excluding below $\bestexclusionHiggsinoThreeNs$ ($\bestexclusionHiggsinoPZeroFiveNs$)\GeV for a lifetime of 3 (0.05)\unit{ns}.
In the wino case, the results of this search are combined with a previous CMS search to produce a result representing the complete LHC data set recorded in 2015--2018, the most stringent constraints to date.
At 95\% confidence level, chargino masses in the wino case are excluded below $\bestexclusionWinoThreeNs$ ($\bestexclusionWinoPTwoNs$)\GeV for a lifetime of 3 (0.2)\unit{ns}.
}

\hypersetup{
pdfauthor={CMS Collaboration},
pdftitle={Search for disappearing tracks in proton-proton collisions at sqrt(s) = 13 TeV},
pdfsubject={CMS},
pdfkeywords={CMS, physics, disappearing tracks}}

\maketitle

\section{Introduction}
\label{sec:introduction}

Many beyond-the-standard-model (BSM) scenarios introduce long-lived charged particles that could decay within the volume of the tracking detectors used by the CERN LHC experiments.
If the decay products of such a particle are undetected, either because they have too little momentum to be reconstructed or because they interact only weakly, a ``disappearing track'' signature is produced. 
This signature is identified as an isolated particle track that extends from the interaction region but is missing hits in the outermost region of the tracking detector, and also has little associated energy deposited in the calorimeters and no associated hits in the outer muon detectors.
Because standard model (SM) processes rarely produce this signature, background processes are almost entirely composed of failures of the particle reconstruction or track finding algorithms.

The disappearing track signature arises in a broad range of BSM scenarios~\cite{Chen:1996ap, Ibe:2012hu,Hall:2012zp, Arvanitaki:2012ps, Citron:2012fg,Garny:2017rxs,Wang:2017sxx,Bharucha:2018pfu,Biswas:2018ybc,Belyaev:2018xpf,Borah:2018smz,Belanger:2018sti,Filimonova:2018qdc}.
For example, in anomaly-mediated supersymmetry breaking (AMSB)~\cite{Giudice:1998xp, Randall:1998uk} the particle mass spectrum includes a chargino and neutralino (electroweakinos $\Pchipm$ and $\Pchiz$, respectively) that are nearly degenerate in mass. 
In this scenario, with a chargino-neutralino mass difference of order $100\MeV$, the chargino is long-lived and can reach the CMS tracking detector before decaying to a neutralino and a pion ($\Pchipm{\to}\Pchiz\PGppm$).
The produced pion does not have sufficient momentum to be reconstructed as a track, nor to contribute significantly to the energy associated with the original chargino track.
The neutralino, as the lightest supersymmetric particle (LSP), is stable assuming R-parity conservation and interacts only weakly, leaving no trace in the detector.
Consequently, the decay of an AMSB chargino to a weakly interacting neutralino and an unreconstructed pion would produce the disappearing track signature.

This letter presents a search for disappearing tracks in proton-proton ($\Pp\Pp$) collision data collected at $\sqrt{s} = 13\TeV$ throughout 2017 and 2018, corresponding to an integrated luminosity of $\intLumi$.
The results of this search are presented in terms of chargino masses and lifetimes within the context of AMSB.
The results are also presented more generally in a form that can be used to test any BSM scenario producing the disappearing track signature.
The ATLAS experiment has previously excluded AMSB, with a purely wino LSP, for chargino masses below $460\GeV$ with a lifetime of $0.2\unit{ns}$~\cite{Aaboud:2017mpt}.
The CMS experiment has excluded AMSB chargino masses for a purely wino LSP below $715\GeV$ for a lifetime of $3\unit{ns}$~\cite{Sirunyan:2018ldc}, using the data collected during 2015 and 2016.
This search extends the previous CMS results to encompass the entire available $\sqrt{s} = 13\TeV$ data set, referred to as the Run~2 data set, corresponding to a total integrated luminosity of $\intLumiRunTwo$. Prior to the 2017 data-taking period, a new pixel detector was installed as part of the Phase~1 upgrade~\cite{CMS:2012sda, Veszpremi:2017yvj}. This new detector contains a fourth inner layer at a radius of $2.9\unit{cm}$ from the interaction region. The addition of this new layer enables this search to accept shorter tracks that traverse fewer layers of the tracker, thereby increasing its sensitivity to shorter lifetime particles. The interpretation of the results is extended to include the direct electroweak production of charginos in the case of a purely higgsino LSP.

\section{The CMS detector}
\label{sec:detector}

The central feature of the CMS apparatus is a superconducting solenoid of
6\unit{m} internal diameter. Within the solenoid volume are a silicon pixel and
strip tracker, a lead tungstate crystal electromagnetic calorimeter (ECAL), and
a brass and scintillator hadron calorimeter (HCAL), each composed of a barrel
and two endcap sections. Forward calorimeters extend the pseudorapidity
coverage provided by the barrel and endcap detectors. Muons are measured in
gas-ionization detectors embedded in the steel flux-return yoke outside the
solenoid.

The silicon tracker measures charged particles within the pseudorapidity range $\abs{\eta} < 2.5$.
During the LHC running period when the data used in this analysis were recorded, the silicon tracker consisted of 1856 silicon pixel and 15\,148 silicon strip detector modules.

Events of interest are selected using a two-tiered trigger system~\cite{Khachatryan:2016bia}. The first level (L1), composed of custom hardware processors, uses information from the calorimeters and muon detectors to select events at a rate of around $100\unit{kHz}$ within a fixed time interval of less than $4\mus$. The second level, known as the high-level trigger (HLT), consists of a farm of processors running a version of the full event reconstruction software optimized for fast processing, and reduces the event rate to $\mathcal{O}(1)\unit{kHz}$ before data storage.

A more detailed description of the CMS detector, together with a definition of
the coordinate system used and the relevant kinematic variables, can be found
in Ref.~\cite{Chatrchyan:2008aa}.

\section{Data sets}
\label{sec:datasets}

This search is based on $\Pp\Pp$ collision data recorded by the CMS detector at $\sqrt{s} = 13\TeV$ corresponding to an integrated luminosity of $\intLumiA$~\cite{CMS-PAS-LUM-17-004} and $\intLumiB$~\cite{CMS-PAS-LUM-18-002} from the 2017 and 2018 data-taking periods, respectively.

Simulated signal events are generated at leading order (LO) precision with \PYTHIA 8.240~\cite{Sjostrand:2014zea}, using the NNPDF3.0~LO~\cite{Ball:2014uwa} parton distribution function (PDF) set with the CP5 tune~\cite{CMS-PAS-GEN-17-001} to describe the underlying event.
Supersymmetric particle mass spectra are produced by \ISAJET 7.70~\cite{Paige:2003mg}, for chargino masses in the range 100--1100 (100--900)\GeV in steps of 100\GeV for the wino (higgsino) LSP case.
The ratio of the vacuum expectation values of the two Higgs doublets (\tanb) is fixed to 5, with a positive higgsino mass parameter ($\mu > 0$).
The $\Pchipm$--$\Pchiz$~mass difference has little dependence on \tanb and the sign of $\mu$~\cite{Ibe:2012sx}.
While this mass difference typically determines the chargino's proper decay time (the lifetime in the rest frame, $\tau$), in these simulated signal events $\tau$ is explicitly varied from $6.67\unit{ps}$ to $333\unit{ns}$ (corresponding to a range in $c\tau$ of 0.2--$10\,000\unit{cm}$) in logarithmic steps, to examine sensitivity to a broader range of models.

In the wino LSP case, the chargino branching fraction ($\mathcal{B}$) for $\Pchipm{\to}\allowbreak\Pchiz \Pgppm$ is set to 100\%, and both $\Pchipm \Pchimp$ and $\Pchipm\Pchiz$ production processes are simulated.
In the higgsino LSP case, the second neutralino ($\PchizTwo$) is completely degenerate in mass with $\Pchiz$, having equal production cross sections ($\sigma$) and branching fractions for the $\Pchipm{\to}\allowbreak\PchizOneTwo + X$ decays.
Following Ref.~\cite{Thomas:1998wy}, these are taken to be 95.5\% for $\Pchipm{\to}\allowbreak\PchizOneTwo \Pgppm$, 3\% for $\Pchipm{\to}\allowbreak\PchizOneTwo\Pe\PGn$, and 1.5\% for $\Pchipm{\to}\allowbreak\PchizOneTwo\PGm\PGn$ in the range of chargino masses of interest, and both $\Pchipm\Pchimp$ and $\Pchipm\PchizOneTwo$ production processes are simulated.

Simulated signal events are normalized using cross sections calculated to next-to-leading order plus next-to-leading-logarithmic (NLO+NLL) precision, 
using \textsc{Resummino}~1.0.9~\cite{Fuks:2012qx,Fuks:2013vua} with the
CTEQ6.6~\cite{Nadolsky:2008zw} and MSTW2008nlo90cl~\cite{Martin:2009iq} PDF sets, and the final numbers are calculated using the PDF4LHC recommendations~\cite{Butterworth:2015oua} for the two sets of cross sections.
In the wino case, the ratio of $\Pchipm\Pchiz$ to $\Pchipm\Pchimp$ production is roughly 2:1 for all chargino masses considered.
In the higgsino case, the ratio of $\Pchipm\PchizOneTwo$ to $\Pchipm\Pchimp$ production is roughly 7:2.

As an LO generator, \PYTHIA is known to be deficient in modeling the rate of initial-state radiation (ISR) and the resulting hadronic recoil~\cite{Chatrchyan:2011ne,Chatrchyan:2013xna}, so data-derived corrections for this deficiency are applied as functions of the transverse momentum (\pt) of the electroweakino pair (either $\Pchipm\Pchimp$ or $\Pchipm\PchizOneTwo$).
Similar to the method used in Ref.~\cite{Chatrchyan:2013xna}, the correction factors are derived as the ratio of the \pt of $\PZ{\to}\PGm\Pgm$ candidates in data to simulated \PYTHIA events, under the assumption that the production of ISR in \PZ\ boson and electroweakino pair events are similar, since both are electroweak processes. 
The ISR correction factors typically range between 1.8 and 2.0 in the kinematic region relevant for this search.

Simulated events are generated with a Monte Carlo program incorporating a full model of the CMS detector, based on \GEANTfour~\cite{Agostinelli:2002hh}, and reconstructed with the same software used for collision data.
Simulated minimum bias events are superimposed on the hard interaction to describe the effect of additional inelastic $\Pp\Pp$ interactions within the same or neighboring bunch crossings, known as pileup, and the samples are weighted to match the pileup distribution observed in data.

\section{Event reconstruction and selection}
\label{sec:selection}

A particle-flow (PF) algorithm~\cite{Sirunyan:2017ulk} aims to reconstruct and identify each individual particle in an event with an optimized combination of information from the various elements of the CMS detector. The energy of photons is obtained from the ECAL measurement. The energy of electrons is determined from a combination of the electron momentum at the primary interaction vertex as determined by the tracker, the energy of the corresponding ECAL cluster, and the energy sum of all bremsstrahlung photons spatially compatible with originating from the electron track. The energy of muons is obtained from the curvature of the corresponding track. 
The energy of charged hadrons is determined from a combination of their momentum measured in the tracker and the matching ECAL and HCAL energy deposits, corrected for the response function of the calorimeters to hadronic showers.
Finally, the energy of neutral hadrons is obtained from the corresponding corrected ECAL and HCAL energies.

For each event, hadronic jets are clustered from these reconstructed particles using the infrared- and collinear-safe anti-\kt algorithm~\cite{Cacciari:2008gp, Cacciari:2011ma} with a distance parameter of 0.4. 
Jet momentum is determined as the vector sum of all particle momenta in the jet, and is found from simulation to be, on average, within 5 to 10\% of the true momentum over the entire \pt spectrum and detector acceptance.
Hadronic \PGt lepton decays are reconstructed with the hadron-plus-strips algorithm~\cite{Sirunyan:2018pgf}, which starts from the reconstructed jets.

The missing transverse momentum vector \ptvecmiss is computed as the negative vector sum of the transverse momenta of all the PF candidates in an event~\cite{Sirunyan:2019kia}, and its magnitude is denoted as \ptmiss. 
The \ptvecmiss is modified to account for corrections to the energy scale of the reconstructed jets in the event.
The related vector \metNoMuVec is calculated in the same manner as \ptvecmiss, excepting that the transverse momenta of PF muons are ignored.
The magnitude of \metNoMuVec is denoted by \metNoMu.
Signal events for this search typically have no reconstructed muons, in which case \ptvecmiss and \metNoMuVec are identical.

As tracking information is not available in the L1 trigger, events are collected by several triggers requiring large \ptmiss or \metNoMu, which would be produced in signal events by an ISR jet recoiling against the electroweakino pair.
The L1 triggers require \ptmiss above a threshold that was varied during the data-taking period according to the instantaneous luminosity.
The HLT requires both \ptmiss and \metNoMu with a range of thresholds.
The lowest threshold trigger, designed specially for this search, requires $\ptmiss > 105\GeV$ and an isolated track with $\pt > 50\GeV$ and at least 5 associated tracker hits at the HLT.
The remaining triggers require $\ptmiss$ or $\metNoMu >$ $120\GeV$ and do not have a track requirement.

After the trigger, events selected offline are required to be consistent with the topology of an ISR jet at the HLT, having $\metNoMu > 120\GeV$, and at least one jet with $\pt > 110\GeV$ and $\abs{\eta} < 2.4$. 
To reject events with spurious \ptmiss from mismeasured jets, the difference in the azimuthal angle $\phi$ between the
direction of the highest \pt jet and \ptvecmiss is required to be greater than 0.5 radians.
For events with at least two jets, the maximum difference in $\phi$ between any two jets, $\Delta\phi_{\text{max}}$, is required to be less than 2.5 radians. 
In 2018, a $40^\circ$ section of one end of the hadronic endcap calorimeter (HEM) lost power during the data-taking period.
The 2018 data are therefore separated into two samples, 2018 A and B, corresponding to events before and after this loss of power, with integrated luminosities of 21 and 39 \fbinv, respectively.
Events from the 2018 B period are rejected if the \ptvecmiss points to the affected region, having $-1.6 < \phi(\ptvecmiss) < -0.6$.
This requirement, referred to as the ``HEM veto'', removes 31\% of the events in 2018 B, and leads to a reduction in the signal acceptance of 16\% for this data-taking period, as expected from geometrical considerations and as verified in simulation.
The selection requirements applied to this point define the ``basic selection'', with the resulting sample dominated by $\Wtolnu$ events.

After the basic selection, isolated tracks with $\pt > 55\GeV$ and $\abs{\eta} < 2.1$ are further selected, where the isolation requirement is defined such that the scalar sum of the \pt of all other tracks within $\Delta R = \sqrt{\smash[b]{(\Delta\eta)^2+(\Delta\phi)^2}} < 0.3$ of the candidate track must be less than 5\% of the candidate track's \pt.
Tracks must be separated from jets having $\pt > 30\GeV$ by $\Delta R(\text{track},\text{jet}) > 0.5$. 
Tracks are also required to be associated with the primary $\Pp\Pp$ interaction vertex (PV), the candidate vertex with the largest value of summed physics-object $\pt^{2}$.
The physics objects in this sum are the jets, clustered with the tracks assigned to candidate vertices as inputs, and the associated missing transverse momentum, taken as the negative vector sum of the \pt of those jets.
With respect to the PV, candidate tracks must have a transverse impact parameter ($\abs{d_{0}}$) less than $0.02\unit{cm}$ and a longitudinal impact parameter ($\abs{d_{z}}$) less than $0.50\unit{cm}$. 

Tracks are said to have a missing hit if they are reconstructed as passing through a functional tracker layer, but no hit in that layer is associated with the track.
A missing hit is described as ``inner'' if the missing layer is between the interaction point and the track's innermost hit, ``middle'' if between the track's innermost and outermost hits, and ``outer'' if it is beyond the track's outermost hit.
The track reconstruction algorithm generally allows for some missing hits, to improve efficiency for tracks traversing the entire tracker. However, for shorter tracks this may result in spurious reconstructed tracks, arising not from charged particle trajectories but from pattern recognition errors.
These spurious tracks are one of two sources of backgrounds considered in this search.
This background is reduced by requiring tracks to have no missing inner or middle hits, and at least four hits in the pixel detector.

The other source of background is isolated, high-$\pt$ charged leptons from SM decays of \Wpm\ or \PZ\ bosons, or from virtual photons.
These tracks can seem to disappear if the track reconstruction fails to find all of the associated hits.
Missing outer hits in lepton tracks may occur because of highly energetic bremsstrahlung in the case of electrons, or nuclear interactions with the tracker material in the case of hadronically decaying \PGt leptons (\tauh).
Electrons or \tauh may be associated with little energy deposited in the calorimeters because of nonfunctional or noisy calorimeter channels.
To mitigate this background, tracks are rejected if they are within $\Delta R(\text{track}, \text{lepton}) < 0.15$ of any reconstructed lepton candidate, whether electron, muon, or \tauh.
This requirement is referred to as the ``reconstructed lepton veto''.
To avoid regions of the detector known to have lower efficiency for lepton reconstruction, fiducial criteria are applied to the track selection. 
In the muon system, tracks within regions of incomplete detector coverage, \ie, within $0.15 < \abs{\eta} < 0.35$ and $1.55 < \abs{\eta} < 1.85$, are rejected.
In the ECAL, tracks in the transition region between the barrel and endcap sections at $1.42 < \abs{\eta} < 1.65$ are rejected, as are tracks whose projected entrance into the calorimeter is within $\Delta R < 0.05$ of a nonfunctional or noisy channel.
Because two layers of the pixel tracker were not fully functional in certain data-taking periods, some regions exhibited low efficiency for the requirement of four or more pixel hits, and tracks within these regions are rejected.
These regions correspond to the range $2.7 < \phi < \pi$ for the region $0 < \eta < 1.42$ in the 2017 data set, and to the range $0.4 < \phi < 0.8$ for the same $\eta$ region in the 2018 data set.
Application of this final requirement rejects approximately 20\% of simulated signal tracks.

Additional regions of lower lepton reconstruction efficiency are identified using ``tag-and-probe'' (T\&P) studies~\cite{Khachatryan:2010xn}. Candidate $\Ztoll$ objects are selected in data where the invariant mass of a tag lepton and a probe track is within $10\GeV$ of $m_{\PZ}$, the world-average mass of the \PZ\ boson~\cite{PDG2018}, resulting in a sample of tracks having a high probability of being a lepton without explicitly requiring that they pass the lepton reconstruction. 
The efficiency of the lepton reconstruction is calculated using these probe tracks across the full coverage of the detector, and also for each local $\eta$-$\phi$ region of size $0.1{\times}0.1$. 
Candidate tracks are rejected from the search region if they are within an $\eta$-$\phi$ region in which the local efficiency is less than the overall mean efficiency by at least two standard deviations.
This procedure removes an additional 4\% of simulated signal tracks.

Finally, two criteria define the condition by which a track is considered to have ``disappeared'': (1) the track must have at least three missing outer hits, and (2) the sum of all associated calorimeter energy within $\Delta R < 0.5$ of the track ($\Ecalo$) must be less than $10\GeV$.
From the sample of tracks passing all of the requirements described above, three signal categories are defined depending on the number of tracker layers that have hits associated to the track, $\nlayers$: $\nlayers = 4$, $\nlayers = 5$, and $\nlayers \geq 6$.
At $\eta = 0$ these categories correspond, respectively, to track lengths of approximately 20, 20--30, and $>$30\unit{cm}.
The previous CMS search for disappearing tracks~\cite{Sirunyan:2018ldc} required at least seven hits associated with the selected tracks, which resulted in a sensitivity comparable to that of only the $\nlayers \geq 6$ category in this search.

\section{Background estimation}
\label{sec:backgroundEstimation}

\subsection{Charged leptons}
\label{sec:chargedLeptonBackground}

For tracks from charged, high-$\pt$ leptons (electrons, muons, or \tauh) to be selected by the search criteria, the lepton reconstruction must fail in such a way that a track is still observed but no lepton candidate is produced, resulting in a mismeasurement of the calorimeter energy in the event. For this reconstruction failure to occur, the following conditions must be present:
\noindent \begin{itemize}
\item There is a reconstructed lepton track that is isolated from other tracks and has zero missing inner or middle hits. In addition, there must be no candidate lepton identified near to it, $\Ecalo$ must be less than $10\GeV$, and the track must have at least three missing outer hits.
\item The resulting \metNoMu must be large enough to pass the offline \metNoMu requirement.
\item The resulting \ptmiss and \metNoMu must be large enough to pass to trigger requirement.
\item In the 2018 B data-taking period, the resulting \ptmiss must pass the HEM veto.
\end{itemize}
The background from charged leptons is estimated by calculating the conditional probability of each of these four requirements in the given order, as described below, treating each lepton flavor independently in each of the three signal categories.

\subsubsection{\texorpdfstring{$\Pveto$}{P(veto)}}
The probability of satisfying the first condition, $\Pveto$, is defined as the probability for a lepton candidate to fail to be identified as a lepton.
This is estimated for electrons (muons) using a T\&P study with $\Ztoee$ ($\Ztomumu$) candidates.
Events are selected if they satisfy a single-electron (single-muon) trigger and contain a tag electron (muon) candidate passing tight identification and isolation criteria.
A probe track is required to pass the disappearing track criteria, excepting the reconstructed lepton veto for the flavor under study, the $\Ecalo$ requirement, and the missing outer hits requirement.
The tag lepton and the probed track are required to have opposite-sign electric charges and an invariant mass within $10\GeV$ of $m_{\PZ}$.

To study these probabilities for \tauh, $\Ztotautau$ candidate events are selected in which one $\PGt$ decays via $\PGt{\to}\Pe\Pgn\Pgn$ or $\PGt{\to}\PGm\Pgn\Pgn$, with the electron or muon serving as the tag lepton.
The other \PGt in these events is selected as the probe track and, after applying the reconstructed electron and muon vetoes to it, the result is a sample of tracks dominated by \tauh.
The electron and muon selections are as described above, with two modifications for the case of \tauh.
To reduce contamination from $\Wtolnu$ events, the transverse mass $\mT = \sqrt{\smash[b]{2 \pt^\ell \ptmiss (1 - \cos \Delta \phi)}}$ is required to be less than $40\GeV$, where $\pt^\ell$ is the magnitude of the tag lepton's transverse momentum and $\Delta \phi$ is the difference in $\phi$ between the \ptvec of the tag lepton and the \ptvecmiss.
In addition, because \PGt\ leptons from the \PZ\ decay are not fully reconstructed, the invariant mass of the tag-probe pair is required to be in the range $m_{\PZ} - 50 < M < m_{\PZ} - 15\GeV$.

For each T\&P study of $\Pveto$ (electrons, muons, and \tauh), the number of selected T\&P pairs before and after applying the relevant flavor of the reconstructed lepton veto, the $\Ecalo$ requirement, and the missing outer hits requirement are labeled $\Ntp$ and $\NtpVeto$, respectively.
To subtract non-\PZ\ boson contributions from the opposite-sign T\&P samples, the selections above are repeated but requiring instead that the tag lepton and probe track have the same electric charge, yielding the quantities $\NSStp$ and $\NSStpVeto$.
The probability that a lepton candidate is not explicitly identified as a lepton is then given by:
\begin{linenomath} \begin{equation}
  \Pveto = \frac{\NtpVeto - \NSStpVeto}{\Ntp - \NSStp}.
\end{equation} \end{linenomath}

The results obtained for $\Pveto$ are summarized in Table~\ref{tab:leptonVetoProbabilities}.

\begin{table*}[htbp]
\centering
\topcaption{Summary of estimated values of $\Pveto$. The uncertainties shown represent only the statistical component.}
\renewcommand{\arraystretch}{1.25}
\begin{tabular}{lcccc}
\multirow{2}{*}{Data-taking period} & \multirow{2}{*}{$\nlayers$} & \multicolumn{3}{c}{$\Pveto$} \\
& & Electrons & Muons & \tauh \\ 
\hline
2017 & 4 & $(8.2 \pm 5.2)\times 10^{-4}$ & $(0.0_{-0.0}^{+3.9})\times 10^{-3}$ & $(6.9_{-5.1}^{+8.3})\times 10^{-2}$ \\
& 5 & $(2.2 \pm 0.9)\times 10^{-4}$ & $(3.2 \pm 1.3)\times 10^{-2}$ & $(6.5^{+2.9}_{-2.7})\times 10^{-2}$ \\
& $\geq$6 & $(2.7 \pm 0.5)\times 10^{-5}$ & $(1.2 \pm 0.5)\times 10^{-6}$ & $(1.0 \pm 0.4)\times 10^{-3}$ \\
[\cmsTabSkip]
2018 A & 4 & $(1.3 \pm 0.7)\times 10^{-3}$ & $(1.0 \pm 1.0)\times 10^{-1}$ & $(7.1_{-3.8}^{+5.5})\times 10^{-2}$ \\
& 5 & $(0.9_{-0.9}^{+1.5})\times 10^{-4}$ & $(7.4 \pm 4.2)\times 10^{-2}$ & $(4.4^{+5.5}_{-4.4})\times 10^{-2}$ \\
& $\geq$6 & $(1.6 \pm 0.6)\times 10^{-5}$ & $(1.9 \pm 0.8)\times 10^{-6}$ & $(0.0_{-0.0}^{+7.3})\times 10^{-4}$ \\
[\cmsTabSkip]
2018 B & 4 & $(0.0_{-0.0}^{+1.1})\times 10^{-4}$ & $(4.0_{-4.0}^{+15.0})\times 10^{-2}$ & $(5.6_{-5.0}^{+6.5})\times 10^{-2}$ \\
& 5 & $(1.4 \pm 1.1)\times 10^{-4}$ & $(5.8 \pm 3.8)\times 10^{-2}$ & $(5.1^{+4.5}_{-3.7})\times 10^{-2}$ \\
& $\geq$6 & $(3.3 \pm 0.7)\times 10^{-5}$ & $(1.5 \pm 0.6)\times 10^{-6}$ & $(2.3 \pm 1.0)\times 10^{-3}$ \\
\end{tabular}
\label{tab:leptonVetoProbabilities}
\end{table*}

\subsubsection{\texorpdfstring{$\Poffline$}{P(off)}}
The probability of satisfying the second condition, $\Poffline$, is defined as the conditional probability of a single-lepton event to pass the offline requirements of $\metNoMu > 120\GeV$ and $\abs{\Delta\phi (\text{leading jet}, \allowbreak \, \metNoMuVec)} > 0.5$, given that the lepton candidate is not explicitly identified as a lepton. The latter of these criteria requires the existence of a jet having $\pt > 110\GeV$ and $\abs{\eta} < 2.4$, as is required in the basic selection.
The $\metNoMuVec$ of events with an unidentified lepton is modeled by assuming the lepton contributes no calorimeter energy to the event, replacing $\metNoMu$ with the magnitude of $\metNoMuVec + \vec{\pt}^{\ell}$.
This modification is applied in single-lepton control samples for each flavor, defined as containing data events passing single-lepton triggers and having at least one tag lepton of the appropriate flavor.
In the case of muons, no modification of $\metNoMuVec$ is made as they are already excluded from its calculation.
The quantity $\Poffline$ is estimated for each lepton flavor by counting the fraction of single-lepton control sample events with $\metNoMu > 120\GeV$ and $\dPhiJetMET > 0.5$, after modifying $\metNoMuVec$ in this way.
For electrons and muons, $\Poffline$ is approximately 0.7--0.8, and approximately 0.2 for \tauh.

\subsubsection{\texorpdfstring{$\Ptrigger$}{P(trig)}}
The probability of satisfying the third condition, $\Ptrigger$, is defined as the conditional probability that a single-lepton event passes the trigger requirement, given that the lepton candidate is not identified as a lepton and the event passes the offline requirements of $\metNoMu > 120\GeV$ and $\dPhiJetMET > 0.5$.
In the single-lepton control samples used to measure $\Poffline$, the efficiency of the trigger requirement is calculated as a function of $\metNoMu$.
The trigger efficiency is then multiplied bin-by-bin by the magnitude of $\metNoMuVec + \vec{\pt}^{\ell}$, described above for $\Poffline$.
The fraction of events in this product that survive the requirement of $\metNoMu$ (modified) $> 120\GeV$ is then the estimate of $\Ptrigger$.
The value of $\Ptrigger$ is approximately 0.3--0.6 for all lepton flavors.

\subsubsection{\texorpdfstring{$\Phemveto$}{P(HEM)}}
The probability of satisfying the fourth condition, $\Phemveto$, is defined as the conditional probability that a single-lepton event survives the HEM veto, given that the lepton candidate is not explicitly identified as a lepton and the event passes both the offline and trigger requirements.
This probability is calculated in the sample of events forming the numerator of $\Ptrigger$.
Because the HEM veto is applied only in the 2018 B data set, $\Phemveto$ is fixed to unity in the other data-taking periods.
The value of $\Phemveto$ is approximately 0.8 for all lepton flavors.

\subsubsection{Charged lepton background estimation}
The product of these four conditional probabilities gives the overall probability for an event with a charged lepton to pass the search selection criteria. 
These probabilities are measured separately for each flavor and within each signal category of $\nlayers$.
To normalize these probabilities to form the background estimate, the number of events with a charged lepton of each flavor ($\Nctrll$) is counted by selecting events passing single-lepton triggers and containing a lepton of the appropriate flavor with $\pt > 55\GeV$.
No requirement on the presence of $\metNoMu$ or the reconstruction of jets is made in counting $\Nctrll$, as $\Poffline$ accounts for the probability to pass those criteria. The value of $\Nctrll$ is corrected by the efficiency of the relevant single-lepton trigger, $\epsilon_{\text{trigger}}^{\ell}$, in order to account for any inefficiencies in that trigger.
From the T\&P samples used to study $\Pveto$, $\epsilon_{\text{trigger}}^{\ell}$ is measured as the fraction of probe tracks satisfying the single-lepton trigger requirement of the $\Nctrll$ selection.
The values are observed to be 84\% in the case of electrons, 94\% in the case of muons, and 90\% in the case of \tauh candidates.
The estimated background from charged leptons is calculated using these components as
\begin{linenomath} \begin{equation}
  \Nestl = \frac{\Nctrll}{\epsilon_{\text{trigger}}^{\ell}} \Pveto \Poffline \Ptrigger \Phemveto.
\end{equation} \end{linenomath}

In the case of the $\nlayers = 4$ and $\nlayers = 5$ signal categories, insufficient numbers of events are available for muons in the estimation of $\Phemveto$, and for muons and \tauh in the estimation of both $\Poffline$ and $\Ptrigger$.
Therefore, these quantities are estimated as the average over the inclusive category $\nlayers \geq 4$.
The dependence of these values on $\nlayers$ for electrons is applied as a systematic uncertainty in these cases, described below in Section~\ref{sec:backgroundSystematics}.

\subsection{Spurious tracks}
\label{sec:spuriousTrackBackground}

Because spurious tracks do not represent the trajectory of an actual charged particle, the combination of tracker layers with associated hits is largely random.
The requirement of zero missing inner and middle hits greatly suppresses the probability of selecting a spurious track.

To measure the probability that an event contains a spurious track, two control samples containing $\Ztoee$ and $\Ztomumu$ decays, respectively, are selected as representative samples of SM events.
The signal benchmark chosen does not contain \PZ\ bosons, so any candidate disappearing tracks observed in these control samples can reliably be labeled as a spurious track.
Since spurious tracks generally do not point to the PV, 
the purity of the spurious tracks samples can be enhanced 
by replacing the nominal requirement of $\abs{d_{0}} < 0.02\unit{cm}$ with a  ``sideband'' selection, defined as  $0.05 \leq \abs{d_{0}} < 0.50\unit{cm}$.

To normalize the sideband selection to the search region, the shape of the $d_{0}$ distribution is described with a fit to a Gaussian function with an added constant, for each control sample in the $\nlayers = 4$ category.
The fit is made in the slightly restricted range $0.1 \leq \abs{d_{0}} < 0.5\unit{cm}$ to remove any overlap with the signal region. 
A transfer factor $\zeta$ is then calculated as the ratio of the integral of the fit function in the signal region to that in the sideband.
The value of $\zeta$ derived from the $\nlayers = 4$ category is applied to the $\nlayers = 5$ and $\nlayers \geq 6$ categories because the event counts in these categories are not sufficient to observe a different $d_{0}$ distribution.
Finally, the spurious track background is estimated as the raw probability for a control sample event to contain a sideband disappearing track candidate ($\Pfakeraw$), multiplied by $\zeta$ and normalized to the number of events passing the basic selection ($\NctrlBasic$):
\begin{linenomath} \begin{equation}
  \NestFake = \NctrlBasic \; \zeta \; \Pfakeraw.
\end{equation} \end{linenomath}
This calculation is performed separately for each signal category of $\nlayers$ for both $\Ztoee$ and $\Ztomumu$ control samples, using the $\Ztomumu$ estimate as the central value of the spurious track background estimate.

\section{Systematic uncertainties}
\label{sec:systematics}

\subsection{Systematic uncertainties in the background estimates}
\label{sec:backgroundSystematics}

The lepton background estimates make the assumption that no visible energy is deposited in the calorimeters by leptons that are not explicitly identified. 
This is tested for electrons and \tauh by allowing selected candidates to deposit $10\GeV$ in the calorimeters, the maximal value allowed by the requirement of $\Ecalo < 10\GeV$ for candidate signal tracks. The modified $\metNoMu$ is constructed as before, but now the calculation includes $10\GeV$ in the direction of the lepton momentum. 
This is applied separately for each $\nlayers$ category for electrons, and in the inclusive $\nlayers \geq 4$ category for \tauh because of small sample sizes.
This results in a 13--15\% decrease in the electron background estimate and an 11--25\% decrease in the \tauh background estimate.
These changes are taken as systematic uncertainties.

In the calculation of $\Poffline$, $\Ptrigger$, and $\Phemveto$, the available data in the $\nlayers = 4$ and $\nlayers = 5$ categories do not separately provide robust measurements for the muon and \tauh background estimates.
Therefore we measure the values in the inclusive category $\nlayers \geq 4$ instead.
The effect of this averaging is estimated by comparing values obtained for these quantities in exclusive and inclusive $\nlayers$ categories for the single-electron control sample, where there is adequate data to measure each.
The differences in these values range between 1 and 11\%.
These values are applied as one-sided systematic uncertainties in the estimate of the background contribution from muon and \tauh candidates for the $\nlayers = 4$ and $\nlayers = 5$ categories.

The spurious track background estimate relies on several assumptions.  The first assumption is that the spurious track probability is independent of the underlying physics content of the event.
This is tested by comparing the estimates obtained from the $\Ztoee$ and $\Ztomumu$ control samples.
The differences in the estimates derived from these two control samples range from $0$ to $200\%$, and are taken as systematic uncertainties in the spurious track background estimate. In every case, the statistical uncertainty in the difference is considerably larger than the difference itself.

The second assumption of the spurious track background estimate is that the projection of the $d_{0}$ sideband correctly describes the signal $d_{0}$ region. This assumption is tested by comparing the number of signal-like tracks $(\abs{d_{0}} < 0.02\unit{cm})$ in the $\Ztoee$ and $\Ztomumu$ control samples to the number projected from the sideband.
Within the statistical and fit uncertainties, the projected number of tracks agrees well with the observed signal-like counts, so no systematic uncertainty is applied.

The third assumption of the spurious track background estimate is that it is independent of the definition of the $d_{0}$ sideband. The validity of this assumption is examined by defining nine alternative, disjoint sidebands of width $0.05\unit{cm}$ instead of the single sideband region of width $0.50\unit{cm}$.
The spurious track estimate is determined for each of these. The observed deviations of these estimates are well within statistical fluctuations of the nominal estimate.
Therefore, no systematic uncertainty is introduced to cover these differences.

The uncertainty in $\zeta$ due to the fit procedure is evaluated by varying the fit parameters within $\pm 1$ standard deviation of their statistical uncertainties, and comparing the resulting values of $\zeta$. A variation of $\pm(43$--$52)$\% from the nominal value is found, and this variation is taken as an estimate of the contribution from this source to the overall systematic uncertainty in the spurious track background.

The systematic uncertainties in the background estimates are summarized in Table~\ref{tab:backgroundSystematics}.

\begin{table*}[htbp]
\centering
\topcaption{Summary of the systematic uncertainties in each background estimate. Each value listed represents the average across all data-taking periods. Some uncertainties are single-sided, as indicated, and those given as a dash are negligible.}
\begin{tabular}{llccc}
\multirow{2}{*}{Background} & \multirow{2}{*}{Source} & \multicolumn{3}{c}{Uncertainty} \\
& & $\nlayers = 4$ & $\nlayers = 5$ & $\nlayers \geq 6$ \\
\hline
Spurious tracks & Control sample & $\pm 19$\% & $\pm 29$\% & $\pm 116$\% \\
& $\zeta$ & $\pm 47$\% & $\pm 47$\% & $\pm 47$\% \\
[\cmsTabSkip]
Electrons & Visible calorimeter energy & $\pm 14$\% & $\pm 14$\% & $\pm 13$\% \\
[\cmsTabSkip]
Muons & $\Poffline$ & $+7$\% & $+7$\% & \NA \\
& $\Ptrigger$ & $+8$\% & $+2$\% & \NA \\
[\cmsTabSkip]
\tauh & Visible calorimeter energy & $\pm 19$\% & $\pm 19$\% & $\pm 19$\% \\
& $\Poffline$ & $+7$\% & $+7$\% & \NA \\
& $\Ptrigger$ & $+8$\% & $+2$\% & \NA \\
\end{tabular}
\label{tab:backgroundSystematics}
\end{table*}

\subsection{Systematic uncertainties in signal selection efficiencies}
\label{sec:signalSystematics}

Theoretical uncertainties in the chargino production cross section arise from the choice of factorization and renormalization scales and from uncertainties in the PDFs used. These effects result in an assigned uncertainty in the expected signal yields of 2--9\%, depending on the chargino mass.
A 2.3 (2.5)\% uncertainty in the total integrated luminosity of the 2017~\cite{CMS-PAS-LUM-17-004} (2018~\cite{CMS-PAS-LUM-18-002}) data set is assigned.
Uncertainties in the signal yields due to corrections or scale factors are evaluated by varying each correction by $\pm 1$ standard deviation of their measured uncertainties, and comparing the resulting signal yields to their nominal value.
The corrections considered include the corrections related to the statistical uncertainty in the ISR corrections (12--15\%) and the modeling of pileup (2--5\%), jet energy scale and resolution (0.1--1.6\%), and $\metNoMu$ (0.1--2.3\%), with the values varying depending on the chargino mass and lifetime.
Uncertainties are estimated in the selection criteria of missing inner, middle, and outer hits (0.1--4.6, 2.5--5.2, and ${<}0.3$\%, respectively) by comparing the efficiency of each between data and simulation in a control sample of single-muon events.
The uncertainty in the efficiency of the $\Ecalo$ requirement is taken to be the difference between the efficiencies obtained from data and from simulation in the $\Ztomumu$ control sample (0.4--1.0\%), where the tracks are expected to be predominantly spurious.
The uncertainty in the track reconstruction efficiency is evaluated to be 2.1\% in 2017 data~\cite{CMS-DP-2019-004} and 2.5\% in 2018 data~\cite{CMS-DP-2020-013}.

The efficiency of the reconstructed lepton veto in simulated events depends on the modeling of detector noise, which may produce calorimeter or muon detector hits that result in a lepton candidate and thereby reject the track.
The differences in reconstructed lepton veto efficiencies between data and simulation are studied by estimating the efficiencies relative to tighter lepton criteria, for which detailed scale factors are available, in the sample of events used to measure $\Pveto$ for the electron and muon backgrounds.
Differences between estimates from data and simulation of up to 0.1\% are observed, and these are taken into account as systematic uncertainties.

Statistical uncertainties in trigger efficiencies for data and simulation are estimated to be 0.4\% for each $\nlayers$ category, and are applied as systematic uncertainties.
In the case of short tracks ($\nlayers = 4$ and $\nlayers = 5$), no source in data is available outside of the search region to measure the efficiency of the track leg of the trigger requirement, which requires at least five tracker hits associated with the track at HLT.
To study this requirement's effect, the trigger efficiency is measured for signal events in each search category as a function of $\metNoMu$, and the differences between $\nlayers \geq 6$ and $\nlayers = 4$ (5) efficiencies are used to define weights for the $\nlayers = 4$ (5) category.
These weights are not applied to the nominal signal yield, but are used to evaluate a conservative systematic uncertainty.
The weighted signal yields are compared to the nominal, unweighted values, resulting in an average systematic uncertainty of 1.0\% (0.3\%) for the $\nlayers = 4$ (5) category.

The systematic uncertainties in the signal efficiencies are summarized in Table~\ref{tab:signalSystematics}.

\begin{table*}[htb]
\centering
\topcaption{Summary of the systematic uncertainties in the signal efficiencies. Each value listed is the average across all data-taking periods, all chargino masses and lifetimes considered, and wino and higgsino cases. The values given as a dash are negligible.}
\begin{tabular}{lccc}
\multirow{2}{*}{Source} & \multicolumn{3}{c}{Uncertainty} \\
& $\nlayers = 4$ & $\nlayers = 5$ & $\nlayers \geq 6$ \\
\hline
Pileup                                           & 3.0\% & 3.3\% & 2.8\%  \\
ISR                                              & 13\% & 13\% & 13\% \\
Trigger efficiency                               & 1.1\% & 0.8\% & 0.4\% \\
Jet energy scale                                 & 0.6\% & 0.7\% & 1.6\% \\
Jet energy resolution                            & 0.5\% & 0.5\% & 1.3\% \\
$\ptmiss$                                        & 0.3\% & 0.3\% & 0.4\% \\
$\Ecalo$                                         & 0.7\% & 0.7\% & 0.7\% \\
Missing inner hits                               & 2.3\% & 1.0\% & 0.3\% \\
Missing middle hits                              & 3.9\% & 5.1\% & 4.4\% \\
Missing outer hits                               & \NA & \NA & 0.2\% \\
Reconstructed lepton veto efficiency                           & 0.1\% & 0.1\% & \NA \\
Track reconstruction efficiency                  & 2.3\% & 2.3\% & 2.3\% \\
[\cmsTabSkip]
Total                                            &  14\% & 15\% & 14\% \\
\end{tabular}
\label{tab:signalSystematics}
\end{table*}

\section{Results}
\label{sec:results}

The expected number of background events and the observed number of events are shown in Table~\ref{tab:results} for each event category and each data-taking period.
The observations are consistent with the expected total background.
Upper limits are set at 95\% confidence level (\CL) on the product of the cross section and branching fraction for each signal model.
These limits are calculated with an asymptotic \CLs criterion~\cite{Junk:1999kv,Read:2002hq,Cowan:2010js}
that uses a test statistic based on a profile likelihood ratio and treats nuisance parameters in a frequentist context.
Nuisance parameters for the theoretical uncertainties in the signal cross sections, integrated luminosity, and signal selection efficiencies are constrained with log-normal distributions.
The uncertainties in the background estimates are estimated separately for spurious tracks and for reconstruction failures of each flavor of charged leptons, and are treated as independent nuisance parameters.
Uncertainties resulting from limited control sample sizes are constrained with gamma distributions, whereas those associated with multiplicative factors or discussed in Section~\ref{sec:backgroundSystematics} are constrained with log-normal distributions.
The three $\nlayers$ categories are treated as independent counting experiments, as are the data-taking periods 2017, 2018 A, and 2018 B.

\begin{table*}[htbp]
\centering
\topcaption{Summary of the estimated backgrounds and the observation. The first and second uncertainties shown are the statistical and systematic contributions, respectively.}
\renewcommand{\arraystretch}{1.25}
\cmsTable{
\begin{tabular}{lccccc}
\multirow{2}{*}{Data-taking period} & \multirow{2}{*}{$\nlayers$} & \multicolumn{3}{c}{Expected backgrounds} & \multirow{2}{*}{Observation} \\
& & Leptons & Spurious tracks & Total & \\
\hline
2017 & 4 & $1.4 \pm 0.9 \pm 0.2$ & $10.9 \pm 0.7 \pm 4.7$ & $12.2 \pm 1.1 \pm 4.7$ & 17 \\
& 5 & $1.1 \pm 0.4 \pm 0.1$ & $1.0 \pm 0.2 \pm 0.6$ & $2.1 \pm 0.4 \pm 0.6$ & 4 \\
& $\geq$6 & $6.7 \pm 1.1 \pm 0.7$ & $0.04 \pm 0.04 _{-0.04}^{+0.08}$ & $6.7 \pm 1.1 \pm 0.7$ & 6 \\
[\cmsTabSkip]
2018 A & 4 & $1.1 _{-0.6}^{+1.0} \pm 0.1$ & $6.2 \pm 0.5 \pm 3.5$ & $7.3 _{-0.8}^{+1.1} \pm 3.5$ & 5 \\
& 5 & $0.2 _{-0.2}^{+0.6} \pm 0.0$ & $0.5 \pm 0.1 \pm 0.3$ & $0.6 _{-0.2}^{+0.6} \pm 0.3$ & 0 \\
& $\geq$6 & $1.8 _{-0.5}^{+0.6} \pm 0.2$ & $0.04 \pm 0.04 _{-0.04}^{+0.06}$ & $1.8 _{-0.5}^{+0.6} \pm 0.2$ & 2 \\
[\cmsTabSkip]
2018 B & 4 & $0.0 _{-0.0}^{+0.8} \pm 0.0$ & $10.3 \pm 0.6 \pm 5.4$ & $10.3 _{-0.6}^{+1.0} \pm 5.4$ & 11 \\
& 5 & $0.4 _{-0.3}^{+0.7} \pm 0.1$ & $0.6 \pm 0.2 \pm 0.3$ & $1.0 _{-0.3}^{+0.7} \pm 0.3$ & 2 \\
& $\geq$6 & $5.7 _{-1.1}^{+1.2} \pm 0.6$ & $0.00 _{-0.00}^{+0.04} \pm 0.00$ & $5.7 _{-1.1}^{+1.2} \pm 0.6$ & 1 \\
\end{tabular}
}
\label{tab:results}
\end{table*}

In the case of electroweak production with a wino LSP, the results of this search are combined with the previous search presented by CMS, based on data collected in 2015 and 2016~\cite{Sirunyan:2018ldc}.
All data-taking periods are treated as completely uncorrelated and are considered as independent counting experiments.
Systematic uncertainties are measured independently for each period and treated as uncorrelated nuisance parameters, with the exception of uncertainties in the signal cross section, which are treated as 100\% correlated.

\begin{figure*}[htbp]
  \centering
  \includegraphics[width=0.49\textwidth]{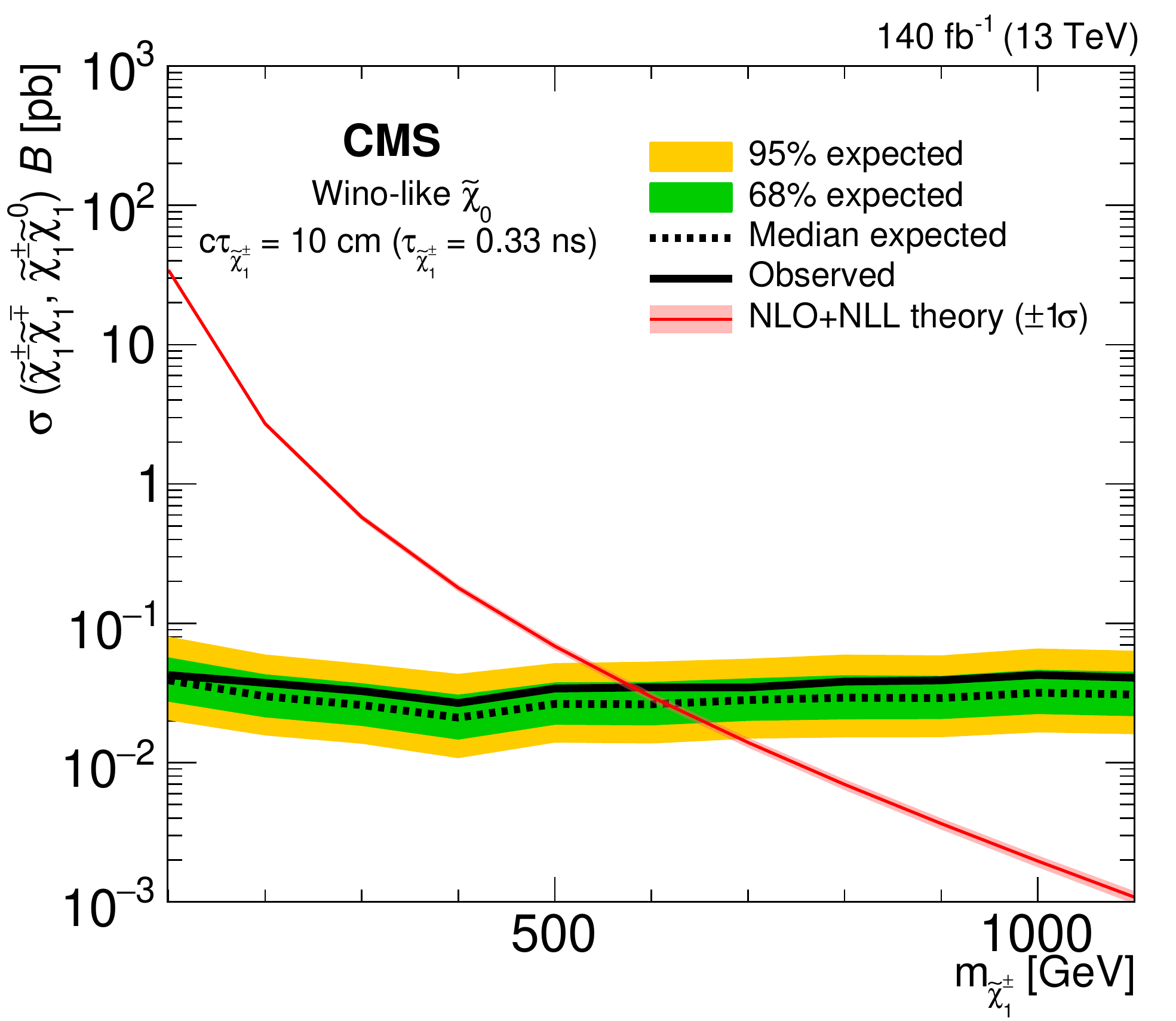}
  \includegraphics[width=0.49\textwidth]{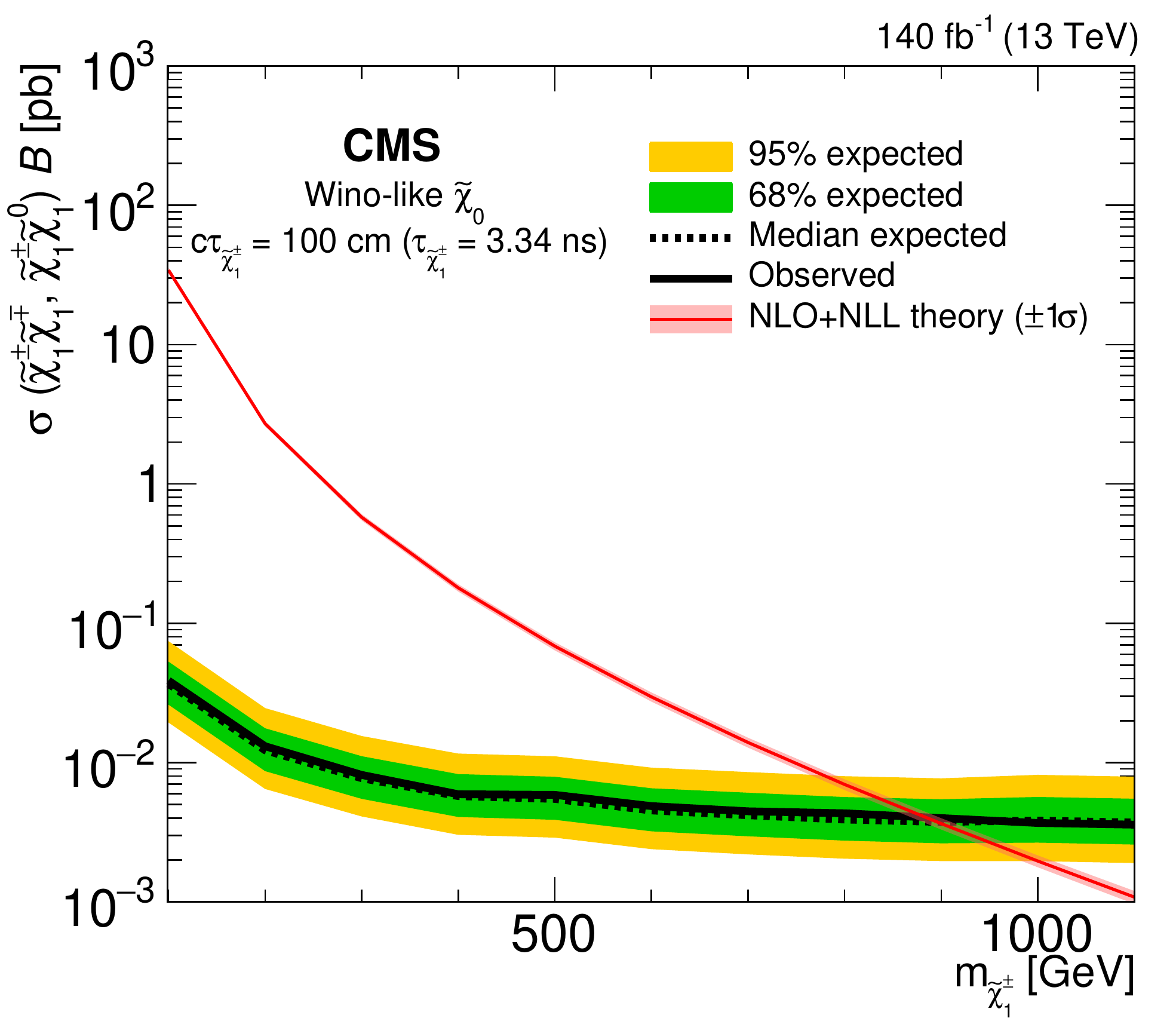}
  \includegraphics[width=0.49\textwidth]{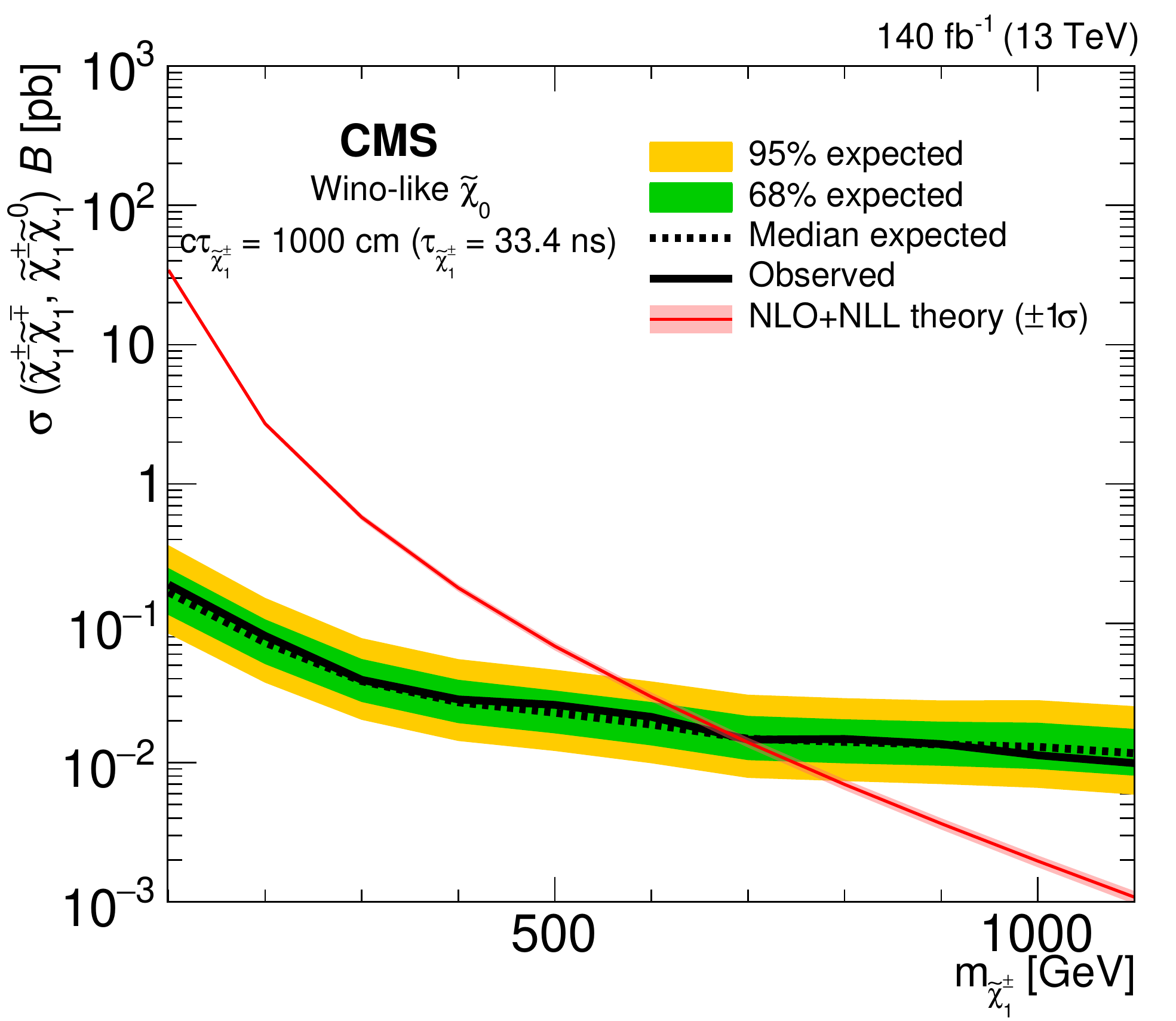}
  \includegraphics[width=0.49\textwidth]{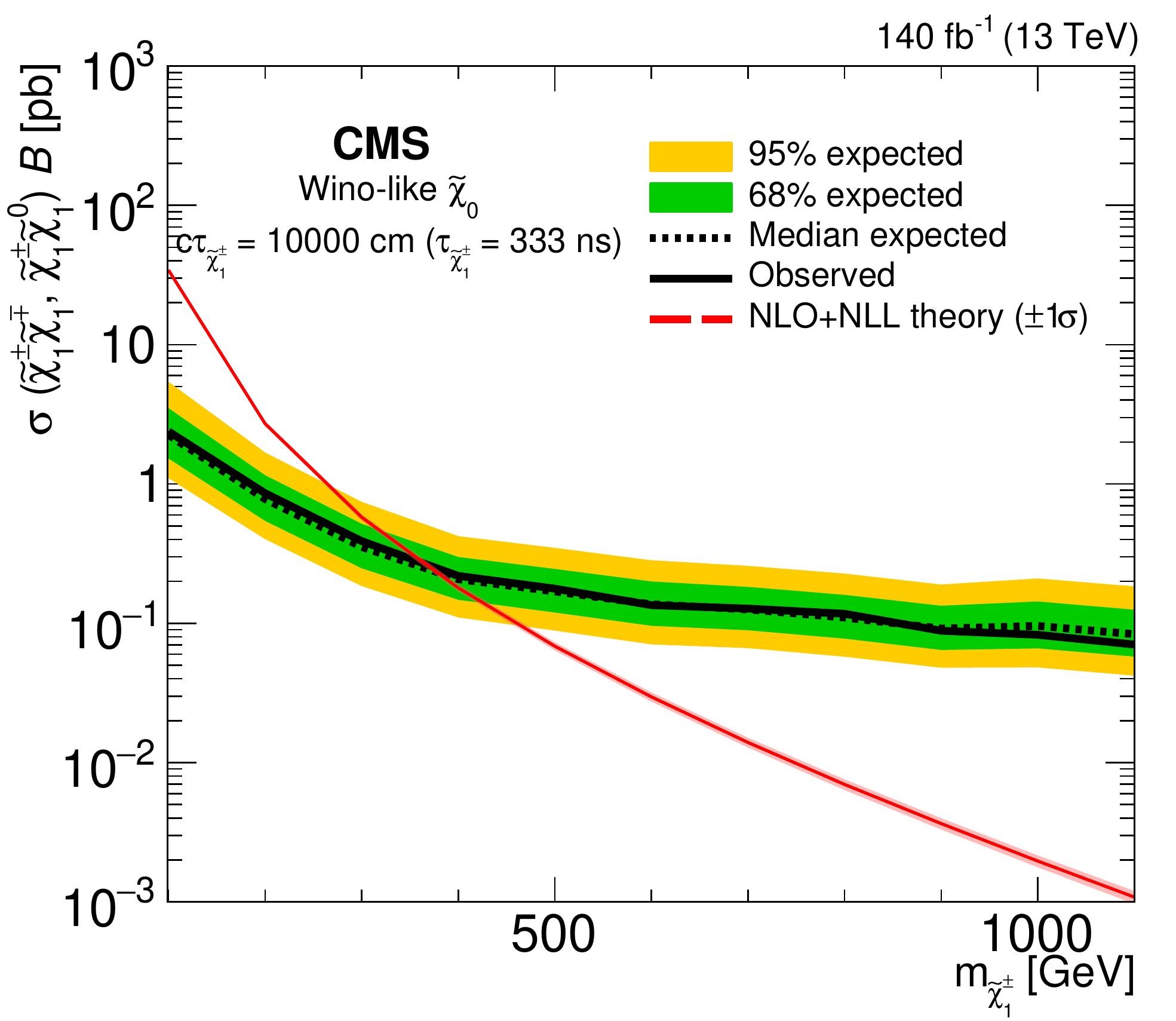}
  \caption{The expected and observed 95\% \CL upper limits on the product of cross section and branching fraction for direct production of charginos as a function of chargino mass for chargino lifetimes of 0.33, 3.34, 33.4, and 333\unit{ns}, for a purely wino LSP with the branching fraction for $\Pchipm{\to}\Pchiz \Pgppm$ set to 100\%.
  Shown are the full Run~2 results, derived from the results of the search in the 2017 and 2018 data sets combined with those of Ref.~\cite{Sirunyan:2018ldc}, obtained in the 2015 and 2016 data sets.
  The cross section includes both $\Pchipm\Pchiz$ and $\Pchipm\Pchimp$ production in roughly a 2:1 ratio for all chargino masses considered.
  The red line indicates the theoretical prediction, described in Section~\ref{sec:datasets}, with scale and PDF uncertainties displayed in the surrounding band.}
  \label{fig:1dWinoLimits}
\end{figure*}

\begin{figure}[htbp]
  \centering
  \includegraphics[width=0.49\textwidth]{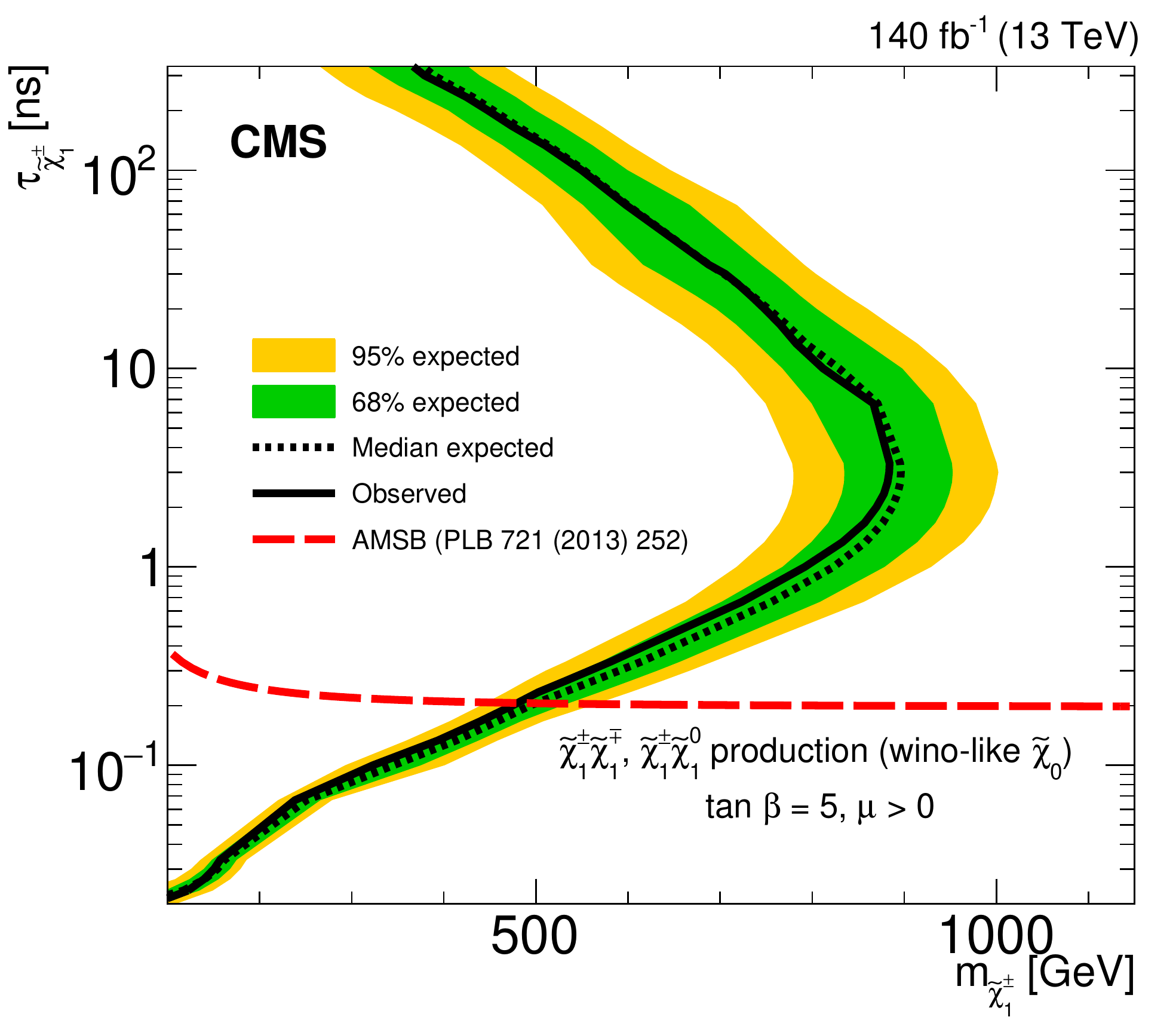} 
  \caption{The expected and observed constraints on chargino lifetime and mass for a purely wino LSP in the context of AMSB, where the chargino lifetime is explicitly varied.
  The chargino branching fraction is set to 100\% for $\Pchipm{\to}\Pchiz \Pgppm$.
  Shown are the full Run~2 results, derived from the results of the search in the 2017 and 2018 data sets combined with those of Ref.~\cite{Sirunyan:2018ldc}, obtained in the 2015 and 2016 data sets.
  The region to the left of the curve is excluded at 95\% \CL.
  The prediction for the chargino lifetime from Ref.~\cite{Ibe:2012sx} is indicated as the dashed line.}
  \label{fig:winoLimits}
\end{figure}

\begin{figure}[htbp]
  \centering
  \includegraphics[width=0.49\textwidth]{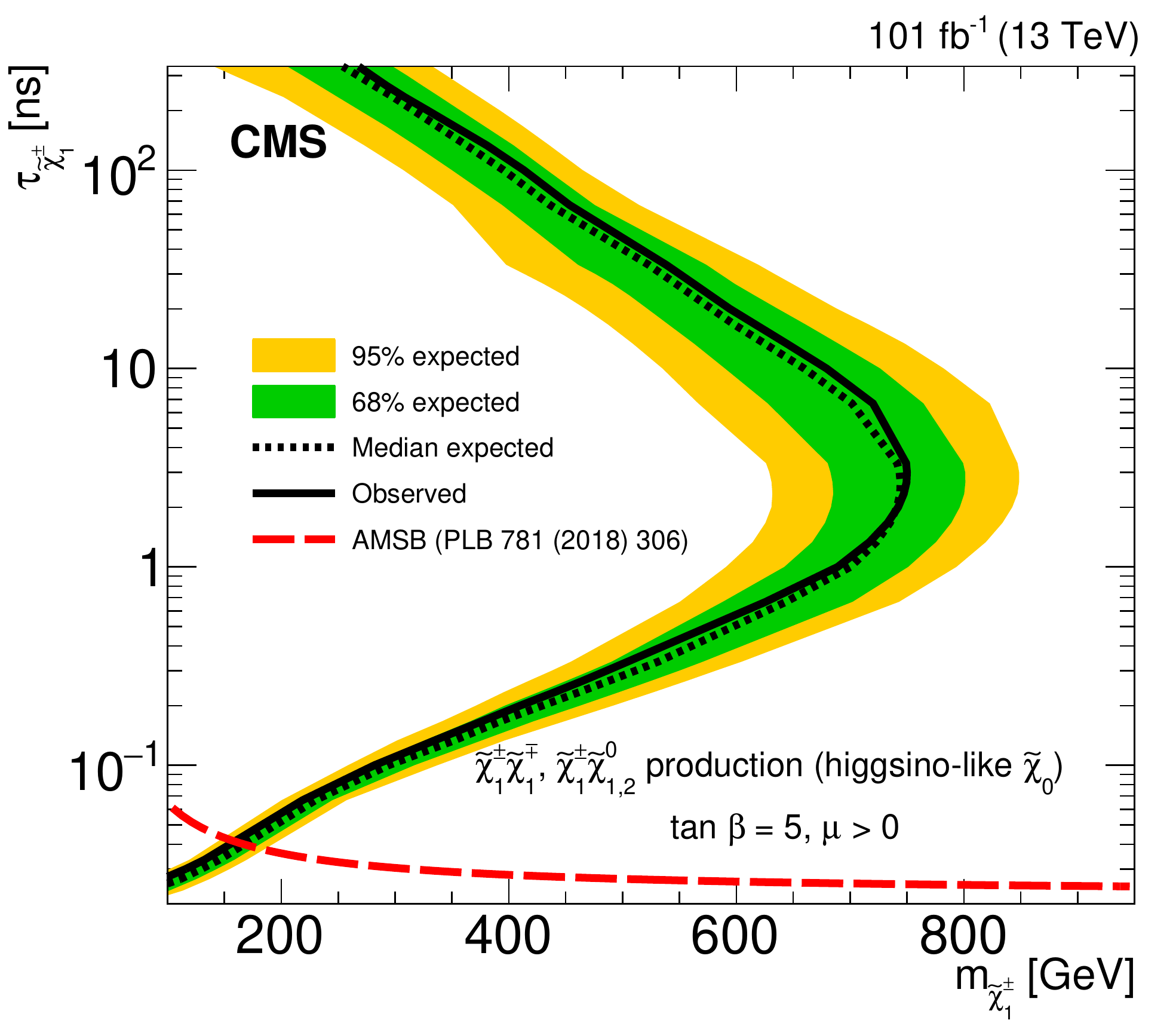}
  \caption{The expected and observed constraints on chargino lifetime and mass for a purely higgsino LSP in the context of AMSB, where the chargino lifetime is explicitly varied.
  Following Ref.~\cite{Thomas:1998wy}, the branching fractions are taken to be 95.5\% for $\Pchipm{\to}\PchizOneTwo\Pgppm$, 3\% for $\Pchipm{\to}\PchizOneTwo\Pe\PGn$, and 1.5\% for $\Pchipm{\to}\PchizOneTwo\PGm\PGn$ in the range of chargino masses of interest, with equal branching fractions and production cross sections between $\Pchiz$ and $\PchizTwo$.
  The region to the left of the curve is excluded at 95\% \CL.
  The prediction for the chargino lifetime from Ref.~\cite{Fukuda:2017jmk} is indicated as the dashed line.}
  \label{fig:higgsinoLimits}
\end{figure}

The expected and observed upper limits on the product of cross sections of electroweak production and branching fractions in the wino LSP case are shown in Fig.~\ref{fig:1dWinoLimits} for four chargino lifetimes.
Two-dimensional constraints derived from the intersection of the theoretical predictions with the expected and observed upper limits, for each chargino mass and mean proper lifetime considered, are shown in Fig.~\ref{fig:winoLimits} for a purely wino LSP and in Fig.~\ref{fig:higgsinoLimits} for a purely higgsino LSP.

Charginos in the wino LSP case with a lifetime of 3 (0.2)\unit{ns} are excluded up to a mass of $\bestexclusionWinoThreeNs$ ($\bestexclusionWinoPTwoNs$)\GeV at 95\% \CL, the most stringent constraints to date.
In the higgsino LSP case, charginos with a lifetime of 3 (0.05)\unit{ns} are excluded up to a mass of $\bestexclusionHiggsinoThreeNs$ ($\bestexclusionHiggsinoPZeroFiveNs$)\GeV at 95\% \CL.
This result is the first to constrain chargino masses with a higgsino LSP obtained with the disappearing track signature.

\section{Summary}
\label{sec:summary}

A search has been presented for long-lived charged particles that decay within the CMS detector and produce a ``disappearing track'' signature.
In the sample of proton-proton collisions recorded by CMS in 2017 and 2018, corresponding to an integrated luminosity of $\intLumi$, $\observationAB$ events are observed, which is consistent with the expected background of $\backgroundAB$ events.
These results are applicable to any beyond-the-standard-model scenario capable of producing this signature and, in combination with the previous CMS search~\cite{Sirunyan:2018ldc}, are the first such results on the complete Run~2 data set, corresponding to a total integrated luminosity of $\intLumiRunTwo$.

Two interpretations of these results are provided in the context of anomaly-mediated supersymmetry breaking.
In the case of a purely higgsino neutralino, charginos are excluded up to a mass of $\bestexclusionHiggsinoThreeNs$ ($\bestexclusionHiggsinoPZeroFiveNs$)\GeV for a mean proper lifetime of 3 (0.05)\unit{ns}, using the 2017 and 2018 data sets.
In the case of a purely wino neutralino, charginos are excluded up to a mass of $\bestexclusionWinoThreeNs$ ($\bestexclusionWinoPTwoNs$)\GeV for a mean proper lifetime of 3 (0.2)\unit{ns}.
These results make use of the upgraded CMS pixel detector to greatly improve sensitivity to shorter particle lifetimes.
For chargino lifetimes above approximately 0.1\unit{ns}, this search places the most stringent constraints on direct chargino production with a purely wino neutralino obtained with the disappearing track signature.
For a purely higgsino neutralino, these constraints are the first obtained with this signature.

\begin{acknowledgments}
  We congratulate our colleagues in the CERN accelerator departments for the excellent performance of the LHC and thank the technical and administrative staffs at CERN and at other CMS institutes for their contributions to the success of the CMS effort. In addition, we gratefully acknowledge the computing centers and personnel of the Worldwide LHC Computing Grid for delivering so effectively the computing infrastructure essential to our analyses. Finally, we acknowledge the enduring support for the construction and operation of the LHC and the CMS detector provided by the following funding agencies: BMBWF and FWF (Austria); FNRS and FWO (Belgium); CNPq, CAPES, FAPERJ, FAPERGS, and FAPESP (Brazil); MES (Bulgaria); CERN; CAS, MoST, and NSFC (China); COLCIENCIAS (Colombia); MSES and CSF (Croatia); RPF (Cyprus); SENESCYT (Ecuador); MoER, ERC IUT, PUT and ERDF (Estonia); Academy of Finland, MEC, and HIP (Finland); CEA and CNRS/IN2P3 (France); BMBF, DFG, and HGF (Germany); GSRT (Greece); NKFIA (Hungary); DAE and DST (India); IPM (Iran); SFI (Ireland); INFN (Italy); MSIP and NRF (Republic of Korea); MES (Latvia); LAS (Lithuania); MOE and UM (Malaysia); BUAP, CINVESTAV, CONACYT, LNS, SEP, and UASLP-FAI (Mexico); MOS (Montenegro); MBIE (New Zealand); PAEC (Pakistan); MSHE and NSC (Poland); FCT (Portugal); JINR (Dubna); MON, RosAtom, RAS, RFBR, and NRC KI (Russia); MESTD (Serbia); SEIDI, CPAN, PCTI, and FEDER (Spain); MOSTR (Sri Lanka); Swiss Funding Agencies (Switzerland); MST (Taipei); ThEPCenter, IPST, STAR, and NSTDA (Thailand); TUBITAK and TAEK (Turkey); NASU (Ukraine); STFC (United Kingdom); DOE and NSF (USA). 
  
  \hyphenation{Rachada-pisek} Individuals have received support from the Marie-Curie program and the European Research Council and Horizon 2020 Grant, contract Nos.\ 675440, 752730, and 765710 (European Union); the Leventis Foundation; the A.P.\ Sloan Foundation; the Alexander von Humboldt Foundation; the Belgian Federal Science Policy Office; the Fonds pour la Formation \`a la Recherche dans l'Industrie et dans l'Agriculture (FRIA-Belgium); the Agentschap voor Innovatie door Wetenschap en Technologie (IWT-Belgium); the F.R.S.-FNRS and FWO (Belgium) under the ``Excellence of Science -- EOS" -- be.h project n.\ 30820817; the Beijing Municipal Science \& Technology Commission, No. Z191100007219010; the Ministry of Education, Youth and Sports (MEYS) of the Czech Republic; the Deutsche Forschungsgemeinschaft (DFG) under Germany's Excellence Strategy -- EXC 2121 ``Quantum Universe" -- 390833306; the Lend\"ulet (``Momentum") Program and the J\'anos Bolyai Research Scholarship of the Hungarian Academy of Sciences, the New National Excellence Program \'UNKP, the NKFIA research grants 123842, 123959, 124845, 124850, 125105, 128713, 128786, and 129058 (Hungary); the Council of Science and Industrial Research, India; the HOMING PLUS program of the Foundation for Polish Science, cofinanced from European Union, Regional Development Fund, the Mobility Plus program of the Ministry of Science and Higher Education, the National Science Center (Poland), contracts Harmonia 2014/14/M/ST2/00428, Opus 2014/13/B/ST2/02543, 2014/15/B/ST2/03998, and 2015/19/B/ST2/02861, Sonata-bis 2012/07/E/ST2/01406; the National Priorities Research Program by Qatar National Research Fund; the Ministry of Science and Education, grant no. 14.W03.31.0026 (Russia); the Tomsk Polytechnic University Competitiveness Enhancement Program and ``Nauka" Project FSWW-2020-0008 (Russia); the Programa Estatal de Fomento de la Investigaci{\'o}n Cient{\'i}fica y T{\'e}cnica de Excelencia Mar\'{\i}a de Maeztu, grant MDM-2015-0509 and the Programa Severo Ochoa del Principado de Asturias; the Thalis and Aristeia programs cofinanced by EU-ESF and the Greek NSRF; the Rachadapisek Sompot Fund for Postdoctoral Fellowship, Chulalongkorn University and the Chulalongkorn Academic into Its 2nd Century Project Advancement Project (Thailand); the Kavli Foundation; the Nvidia Corporation; the SuperMicro Corporation; the Welch Foundation, contract C-1845; and the Weston Havens Foundation (USA). 
\end{acknowledgments}
\ifthenelse{\boolean{cms@external}}{\clearpage}{}
\bibliography{auto_generated}

\cleardoublepage \appendix\section{The CMS Collaboration \label{app:collab}}\begin{sloppypar}\hyphenpenalty=5000\widowpenalty=500\clubpenalty=5000\vskip\cmsinstskip
\textbf{Yerevan Physics Institute, Yerevan, Armenia}\\*[0pt]
A.M.~Sirunyan$^{\textrm{\dag}}$, A.~Tumasyan
\vskip\cmsinstskip
\textbf{Institut f\"{u}r Hochenergiephysik, Wien, Austria}\\*[0pt]
W.~Adam, F.~Ambrogi, T.~Bergauer, M.~Dragicevic, J.~Er\"{o}, A.~Escalante~Del~Valle, R.~Fr\"{u}hwirth\cmsAuthorMark{1}, M.~Jeitler\cmsAuthorMark{1}, N.~Krammer, L.~Lechner, D.~Liko, T.~Madlener, I.~Mikulec, N.~Rad, J.~Schieck\cmsAuthorMark{1}, R.~Sch\"{o}fbeck, M.~Spanring, S.~Templ, W.~Waltenberger, C.-E.~Wulz\cmsAuthorMark{1}, M.~Zarucki
% ...
\vskip\cmsinstskip
\dag: Deceased\\
1:  Also at Vienna University of Technology, Vienna, Austria\\
% ...
\end{sloppypar}
\end{document}